\documentclass[twocolumn,aps,prc,a4paper,superscriptaddress,nofootinbib]{revtex4-1}
\usepackage[utf8]{inputenc}
\usepackage{amsmath}
\usepackage{amssymb}
\usepackage{graphicx}
\usepackage{esint}
\usepackage[colorlinks=true, linkcolor=blue, citecolor=red, urlcolor=black]{hyperref}

\begin{document}

\title{Global $\Lambda$ polarization in heavy-ion collisions from a transport
model}

\author{Hui Li}
\email{lihui12@mail.ustc.edu.cn}
\affiliation{Department of Modern Physics, University of Science and Technology
of China, Hefei, Anhui 230026, China}

\author{Long-Gang Pang}
\email{pang@fias.uni-frankfurt.de}
\affiliation{Frankfurt Institute for Advanced Studies, Ruth-Moufang-Strasse 1,
60438 Frankfurt am Main, Germany}

\author{Qun Wang}
\email{qunwang@ustc.edu.cn}
\affiliation{Department of Modern Physics, University of Science and Technology
of China, Hefei, Anhui 230026, China}

\author{Xiao-Liang Xia}
\email{xiaxl@mail.ustc.edu.cn}
\affiliation{Department of Modern Physics, University of Science and Technology
of China, Hefei, Anhui 230026, China}

\begin{abstract}
The polarizations of $\Lambda$ and $\bar{\Lambda}$ hyperons are
important quantities in extracting the fluid vorticity of the strongly
coupled quark gluon plasma and the magnitude of the magnetic field
created in off-central heavy-ion collisions, through the spin-vorticity
and spin-magnetic coupling. We computed the energy dependence of the
global $\Lambda$ polarization in off-central Au+Au collisions in
the energy range $\sqrt{s_{NN}}=7.7-200$ GeV using a multiphase transport
model. The observed polarizations with two different impact parameters
agree quantitatively with recent STAR measurements. The energy dependence
of the global $\Lambda$ polarization is decomposed as energy dependence
of the $\Lambda$ distribution at hadronization and the space-time
distribution of the fluid-vorticity field. The visualization of both
the $\Lambda$ distribution and the fluid-vorticity field show a smaller
tilt at higher collisional energies, which indicates that the smaller
global polarization at higher collisional energies is caused by a
smaller angular momentum deposition at midrapidity.
\end{abstract}
\maketitle

\section{Introduction}

In off-central heavy-ion collisions, huge orbital angular momenta
of order $10^{3}-10^{5}\hbar$ are generated. How such orbital angular
momenta are distributed in the hot and dense matter is an interesting
topic to be investigated. There is an inherent correlation between
rotation and particle polarization. The Einstein-de Haas effect \cite{einstein1915}
demonstrates that a sudden magnetization of the electron spins in
a ferromagnetic material leads to a mechanical rotation due to angular
momentum conservation. Barnett \cite{barnett1915} proved the existence
of the reverse process – the rotation of an uncharged body leads to
the polarization of atoms and spontaneous magnetization. It is expected
that quarks are also polarized in the rotating quark-gluon plasma
(QGP) created in off-central heavy-ion collisions. Liang and Wang
first proposed that $\Lambda$ hyperons can be polarized along the
orbital angular momentum of two colliding nucleus \cite{Liang:2004ph,Gao:2007bc}.
Voloshin suggested that such a polarization can even be observed in
unpolarized hadron-hadron collisions \cite{Voloshin:2004ha}. Besides
the global orbital angular momentum, the local vorticity created by
a fast jet going through the QGP also affects the hadron polarization
\cite{Betz:2007kg}. The polarization density near equilibrium is
first computed in the statistical-hydrodynamic model \cite{Becattini:2007sr,Becattini:2007nd,Becattini:2013fla}
and later confirmed in a quantum kinetic approach \cite{Fang:2016vpj}.
Some hydrodynamic calculations quantitatively predicted the global
polarization in off-central heavy-ion collisions \cite{Becattini:2013vja,Karpenko:2016jyx,Xie:2016fjj,Xie:2017upb}.
The fluid vorticity has also been investigated in transport simulations
\cite{Jiang:2016woz,Deng:2016gyh}. For more studies of the fluid
vorticity and $\Lambda$ polarization, please refer to Refs. \cite{Csernai:2013bqa,Csernai:2014ywa,Becattini:2015ska,Pang:2016igs,Aristova:2016wxe,Baznat:2013zx,Teryaev:2015gxa,Ivanov:2017dff}.

Recently STAR measured the global polarization of $\Lambda$ and $\bar{\Lambda}$
in off-central Au+Au collisions in the Beam Energy Scan (BES) program
\cite{STAR:2017ckg}. From the measured polarization, the fluid vorticity
of the strongly coupled QGP and the magnitude of the magnetic field
created in off-central heavy-ion collisions are extracted for the
first time using the spin-vorticity and spin-magnetic coupling. It
indicates that the rotational fluid has the largest vorticity, of
the order of $10^{21}\ \text{s}^{-1}$, that ever existed in the universe.
So the strongly coupled QGP has an additional extreme feature: it
is the fluid with the highest vorticity. The global polarization of
hyperons plays an important role in probing the vorticity field of
the QGP. Therefore, it is worth to study the inherent correlation
between the global polarization and the microscopic vortical structure
in detail.

In this paper, we focus on the energy dependence of the vorticity
field and global $\Lambda$ polarization within a multiphase transport
(AMPT) model \cite{Lin:2004en} for nuclear-nuclear collisions in
the energy range $\sqrt{s_{NN}}=7.7-200$ GeV. The vorticity field
profile given by AMPT is used to compute the global polarization of
$\Lambda$ and $\bar{\Lambda}$ produced in the hadronization stage
using the spin-vorticity coupling. The paper is organized as follows.
In Sec.~\ref{sec:Polarization-lambda}, we give the formula for the
polarization induced by vorticity. In Sec.~\ref{sec:setup}, we introduce
the numerical method we use. The numerical results and discussions
are presented in Sec.~\ref{sec:result}. We finally give a summary
in Sec.~\ref{sec:summary}.

In this paper, we use the following conventions. The metric tensor
is chosen as $g_{\mu\nu}=\mathrm{diag}(1,-1,-1,-1)$ and the Levi-Civita
symbol satisfies $\epsilon^{0123}=1$. The symbols in boldface represent
the spatial components of four-vectors, for example, $S^{\mu}=(S^{0},\mathbf{S})$
denotes the spin four-vector and $u^{\mu}=\gamma(1,\mathbf{v})$ denotes
the fluid velocity four-vector with $\gamma=1/\sqrt{1-\mathbf{v}^{2}}$
being the Lorentz factor.

\section{$\Lambda$ Polarization from vorticity}

\label{sec:Polarization-lambda}In local thermal equilibrium, the
ensemble average of the spin vector for spin-$1/2$ fermions with
four-momentum $p$ at space-time point $x$ is obtained from the statistical-hydrodynamical
model \cite{Becattini:2013fla} as well as the Wigner function approach
\cite{Fang:2016vpj} and reads
\begin{equation}
S^{\mu}(x,p)=-\frac{1}{8m}\left(1-n_{F}\right)\epsilon^{\mu\nu\rho\sigma}p_{\nu}\varpi_{\rho\sigma}(x),\label{eq:spin_thermal}
\end{equation}
where the thermal vorticity tensor is given by
\begin{equation}
\varpi_{\mu\nu}=\frac{1}{2}\left(\partial_{\nu}\beta_{\mu}-\partial_{\mu}\beta_{\nu}\right),\label{eq:thermal_vorticity}
\end{equation}
with $\beta^{\mu}=u^{\mu}/T$ being the inverse-temperature four-velocity.
In Eq.~(\ref{eq:spin_thermal}), $m$ is the mass of the particle
and $n_{F}=1/[1+\exp(\beta\cdot p\mp\mu/T)]$ is the Fermi-Dirac distribution
function for particles ($-$) and anti-particles ($+$).

Some approximations can be made in Eq.~(\ref{eq:spin_thermal}) to
simplify the computation of the global $\Lambda$ polarization. Since
the temperature at hadronization is much lower than the mass of the
$\Lambda$, the number density of $\Lambda$'s is very small so that
we can make the approximation $1-n_{F}\simeq1$ as in Ref. \cite{Becattini:2013vja}.
With this approximation, Eq.~(\ref{eq:spin_thermal}) is the same
for $\Lambda$ and $\bar{\Lambda}$. Care has to be taken since a
finite chemical potential in $n_{F}$ or spin-magnetic coupling could
induce a difference between $\Lambda$ and $\bar{\Lambda}$. However,
the difference between the polarizations of $\Lambda$ and $\bar{\Lambda}$
measured in the STAR experiment is not distinguishable within errors
\cite{STAR:2017ckg,Upsal:2017QM}. For simplicity, we also do not
distinguish $\Lambda$ and $\bar{\Lambda}$ in the present research.
Therefore, Eq.~(\ref{eq:spin_thermal}) is rewritten as
\begin{equation}
S^{\mu}(x,p)=-\frac{1}{8m}\epsilon^{\mu\nu\rho\sigma}p_{\nu}\varpi_{\rho\sigma}(x).\label{eq:spin_relativistic}
\end{equation}
By decomposing the thermal vorticity in Eq.~(\ref{eq:thermal_vorticity})
into the following components,
\begin{align}
\boldsymbol{\varpi}_{T} & =(\varpi_{0x},\varpi_{0y},\varpi_{0z})=\frac{1}{2}\left[\nabla\left(\frac{\gamma}{T}\right)+\partial_{t}\left(\frac{\gamma\mathbf{v}}{T}\right)\right],\nonumber \\
\boldsymbol{\varpi}_{S} & =(\varpi_{yz},\varpi_{zx},\varpi_{xy})=\frac{1}{2}\nabla\times\left(\frac{\gamma\mathbf{v}}{T}\right),\label{eq:vorticity_components}
\end{align}
Eq.~(\ref{eq:spin_relativistic}) can be rewritten as
\begin{align}
S^{0}(x,p) & =\frac{1}{4m}\mathbf{p}\cdot\boldsymbol{\varpi}_{S},\nonumber \\
\mathbf{S}(x,p) & =\frac{1}{4m}\left(E_{p}\boldsymbol{\varpi}_{S}+\mathbf{p}\times\boldsymbol{\varpi}_{T}\right),\label{eq:spin_components}
\end{align}
where $E_{p}$, $\mathbf{p}$, $m$ are the $\Lambda$'s energy, momentum,
and mass, respectively.

The spin vector in Eq.~(\ref{eq:spin_components}) is defined in the
center of mass (c.m.) frame of Au+Au collisions. In the STAR experiment,
the $\Lambda$ polarization is measured in the local rest frame of
the $\Lambda$ by its decay proton's momentum. The spin vector of
$\Lambda$ in its rest frame is denoted as $S^{*\mu}=(0,\mathbf{S}^{*})$
and is related to the same quantity in the c.m. frame by a Lorentz
boost
\begin{equation}
\mathbf{S}^{*}(x,p)=\mathbf{S}-\frac{\mathbf{p}\cdot\mathbf{S}}{E_{p}\left(m+E_{p}\right)}\mathbf{p}.\label{eq:spin_rest_frame}
\end{equation}
By taking the average of $\mathbf{S}^{*}$ over all $\Lambda$ particles
produced at the hadronization stage of AMPT, we obtain the average
spin vector
\begin{equation}
\left\langle \mathbf{S}^{*}\right\rangle =\frac{1}{N}\sum_{i=1}^{N}\mathbf{S}^{*}(x_{i},p_{i}),\label{eq:spin_average}
\end{equation}
where $N$ is the number of $\Lambda$s in all events and $i$ labels
one individual $\Lambda$. The global $\Lambda$ polarization in the
STAR experiment is the projection of $\left\langle \mathbf{S}^{*}\right\rangle $
onto the direction of global angular momentum in off-central collisions
(normal to the reaction plane),
\begin{equation}
P=2\frac{\left\langle \mathbf{S}^{*}\right\rangle \cdot\mathbf{J}}{|\mathbf{J}|},\label{eq:polarization}
\end{equation}
where we have included a normalization factor ($P$ is normalized
to 1) and $\mathbf{J}$ denotes the global orbital angular momentum
of off-central collisions.

\section{Model setup}

\label{sec:setup}
\begin{figure}
\begin{centering}
\includegraphics[width=0.8\columnwidth]{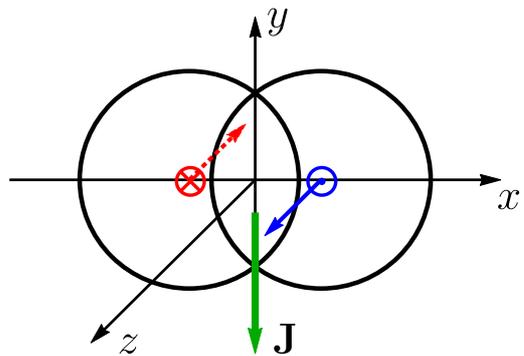}
\par\end{centering}
\caption{\label{fig:geometrical illustration} Schematic picture of an off-central
collision.}
\end{figure}

The string-melting version of the AMPT model is
employed as event generator. It contains four stages: the initial
condition, a parton cascade, hadronization, and hadronic rescatterings.
In this paper, we coarse-grain the parton stage to calculate the thermal
vorticity, and collect $\Lambda$ hyperons produced in the hadronization
stage for calculating the global polarization. Some notations are
made as follows. The reaction plane is fixed to be the $x$-$z$ plane
where $x$ is the direction of impact parameter $\mathbf{b}$ and
$z$ is the beam direction as shown in Fig.~\ref{fig:geometrical illustration}.
In off-central collisions, one nucleus centered at $(x=b/2,\ y=0)$
in the transverse plane moves along the $z$ direction, while the
other nucleus centered at $(x=-b/2,\ y=0)$ moves along the $-z$
direction, with $b\equiv|\mathbf{b}|$. The total angular momentum
$\mathbf{J}$ of the system and the average spin vector $\left\langle \mathbf{S}^{*}\right\rangle $
thus point along the $-y$ direction. However, the $y$-component
of the local thermal vorticity $\boldsymbol{\varpi}_{S}$ ($\varpi_{zx}$)
is not forced to be negative everywhere in the fireball. In fact the
local vorticity is a measure of local rotation in the comoving frame
of one cell. Both relativistic fluid dynamics and transport models
exhibit rich local vorticity structures \cite{Pang:2016igs,Deng:2016gyh,Jiang:2016woz,Csernai:2013bqa,Csernai:2014ywa,Ivanov:2017dff,Teryaev:2015gxa,Becattini:2015ska}.
The global angular momentum is the integral of local vorticity over
all regions.

The AMPT model tracks the positions and momenta of all particles at
any given time. These particles need to be fluidized on space-time
grids in order to calculate the velocity field numerically \cite{Oliinychenko:2015lva,Deng:2016gyh,Jiang:2016woz}.
In the present study, the space-time volume of the system is divided
into 30 time steps with the interval $\Delta t=1\ \text{fm}/c$, $41\times41$
cells in transverse plane with spacing $\Delta x=\Delta y=0.5$ fm,
and 21 cells in the rapidity direction of size $\Delta\eta=0.5$ unit.
Each cell is labeled by its time $t$ and the coordinate of its center
$(x,y,z)$. The thermal vorticity $\varpi_{\mu\nu}$ in each space-time
cell is constructed by the following method. We first calculate the
energy-momentum tensor $T^{\mu\nu}$ in each cell by computing the
sum of $p^{\mu}p^{\nu}/E$ of all particles in the cell and taking
an average over many events,
\begin{equation}
T^{\mu\nu}\left(t,x,y,z\right)=\frac{1}{N_{e}\Delta V}\underset{i}{\sum}\underset{j}{\sum}\frac{p_{ij}^{\mu}p_{ij}^{\nu}}{E_{ij}},\label{eq:T_mu_nu}
\end{equation}
where $p_{ij}^{\mu}=(E_{ij},\mathbf{p}_{ij})$ denotes the $j$-th
particle's four-momentum in a certain cell in the $i$-th event, $\Delta V$
represents the volume of the cell and $N_{e}$ is the number of events.
Then the four-velocity $u^{\mu}$ as well as the energy density $\varepsilon$
in each cell are obtained by solving the eigenvalue problem $T_{\phantom{\mu}\nu}^{\mu}u^{\nu}=\varepsilon u^{\mu}$,
where $u^{\mu}$ is normalized by $u^{\mu}u_{\mu}=1$. The temperature
field $T$ is determined from $\varepsilon$ using the equation of
state $\varepsilon(T)$ in Lattice QCD \cite{Borsanyi:2012cr,Bazavov:2017dus}.
Finally, the obtained velocity field and temperature field are used
to calculate the thermal vorticity for each cell following Eq.~(\ref{eq:thermal_vorticity}),
using the finite-difference method (FDM).

For each BES energy and our chosen impact parameter (7 and 9 fm),
we generate $10^{5}$ events and take the event average in Eq.~(\ref{eq:T_mu_nu})
over them. In this way, the event-by-event fluctuation of the velocity
is removed, so some event-by-event structures of the fluid, such as
vortex pairings in the transverse plane due to hot spots \cite{Pang:2016igs}
are wiped out. However, this is not a problem for the current study
on the global $\Lambda$ polarization since it is an integral effect
of all regions and the local fluctuations are canceled or smeared
when taking the sum over contributions.

Once the thermal vorticity field is stored in the four-dimensional
space-time cells, the spin vector of a $\Lambda$ hyperon can be computed
from the spin-vorticity coupling as given in Eq.~(\ref{eq:spin_components}),
using the value of the local thermal vorticity in the cell where and
when the $\Lambda$ is produced. Then the global $\Lambda$ polarization
is obtained by taking average over the spin vectors of all $\Lambda$
hyperons by Eq.~(\ref{eq:spin_average}) and is projected on the angular
momentum direction by Eq.~(\ref{eq:polarization}). For each chosen
energy and impact parameter, about $2\times10^{6}$ $\Lambda$ hyperons
are used to take the average.

\section{Results and discussions\label{sec:result}}

\subsection{Results for the $\Lambda$ polarization}

\begin{figure}
\begin{centering}
\includegraphics[width=1\columnwidth]{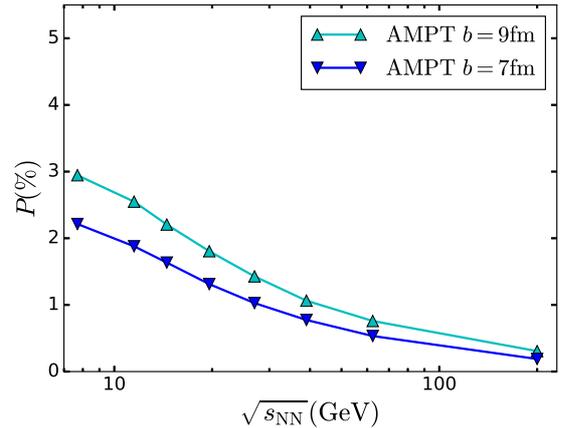}
\par\end{centering}
\caption{\label{fig:Collision-energy-dependence} The global $\Lambda$ polarization
at two impact parameters $b=7$ fm (the blue lower line) and 9
fm (the cyan upper line).}
\end{figure}

We run simulations at BES energies $\sqrt{s_{NN}}=7.7$, 11.5, 14.5,
19.6, 27, 39, 62.4 and 200 GeV. For each energy, we choose two fixed
impact parameters $b=7$ fm and 9 fm from the range $b=5.5-11.3$
fm corresponding to the $20\%-50\%$ centrality class of the STAR
experiment \cite{Qiu:2013wca}. To match the Time Projection Chamber
(TPC) region in the STAR experiment \cite{STAR:2017ckg}, $\Lambda$
hyperons are selected from the midrapidity region $|\eta|<1$. We
calculate the global $\Lambda$ polarization using the method described
in Sec. \ref{sec:setup}. The results are shown in Fig.~\ref{fig:Collision-energy-dependence}.

As shown in Fig.~\ref{fig:Collision-energy-dependence}, the global
polarization is largest at $\sqrt{s_{NN}}=7.7$ GeV and decreases
as the collisional energy increases. It almost vanishes at $\sqrt{s_{NN}}=200$
GeV. The global polarization at $b=9$ fm is larger than that at $b=7$
fm at a specific energy. This is consistent with previous studies
where the averaged and weighted vorticity increases with the impact
parameter in the range $b<10$ fm \cite{Deng:2016gyh,Jiang:2016woz}.
The results shown in Fig.~\ref{fig:Collision-energy-dependence}
only contain primary $\Lambda$ hyperons that are directly produced
at hadronization.

In practice, some $\Lambda$ hyperons are secondary particles produced
from resonance decays, like $\Sigma(1385)\rightarrow\Lambda+\pi$
(strong decay) or $\Sigma^{0}\rightarrow\Lambda+\gamma$ (electromagnetic
decay). It was shown in Ref. \cite{Becattini:2016gvu} that including
feed-down $\Lambda$s decreases the global polarization. Different
decay channels or decay parameters give different suppression factors.
For direct decays and two-step cascade decays, the global polarization
is estimated to be suppressed by about 17\%, using the contributions
to $\Lambda$ from Ref. \cite{Qiu:2013wca} and the decay branching
ratios from Ref. \cite{Agashe:2014kda}. For comparison, the suppression
ratio is estimated to be 15\% in Ref. \cite{Karpenko:2016jyx} and
20\% in Ref. \cite{Becattini:2016gvu}.

\begin{figure}
\begin{centering}
\includegraphics[width=1\columnwidth]{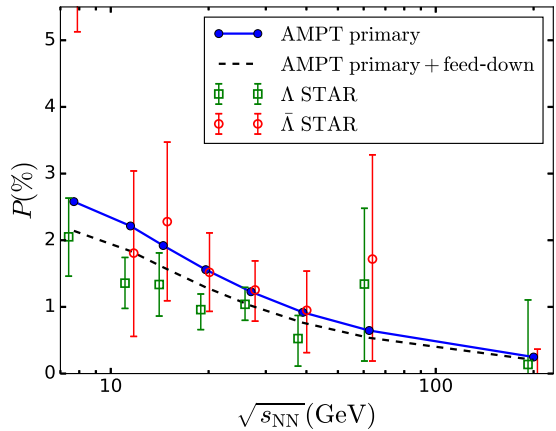}
\par\end{centering}
\caption{\label{fig:Collision-energy-dependence-2} The global $\Lambda$ polarization
at energies from 7.7 GeV to 200 GeV. The blue solid line represents
the global polarization of primary $\Lambda$s and the black dashed
line shows global polarization of primary plus feed-down $\Lambda$s.
The unfilled squares and circles represent the global $\Lambda$ and
$\bar{\Lambda}$ polarization measurements at STAR \cite{Abelev:2007zk,STAR:2017ckg}.}
\end{figure}

In Fig.~\ref{fig:Collision-energy-dependence-2}, we compare our
results with the STAR data. The solid line represents the global polarization
of primary $\Lambda$s from the average over two impact parameters.
Primary plus feed-down $\Lambda$s result in a suppression of 17\%,
as shown by the dashed line, which is closer to the data than the
one for primary $\Lambda$s only. The splitting between $\Lambda$
and $\bar{\Lambda}$ is not included in the current study.

\subsection{Collisional energy dependence of global polarization}

As shown in last subsection, the global $\Lambda$ polarization decreases
as the collisional energy increases: the value of $P$ at 7.7 GeV
is more than 10 times of that at 200 GeV. This behavior contradicts
the energy dependence of the global angular momentum. The reason for
a small global polarization at high collisional energy where angular
momentum is large is investigated in this section.

According to our numerical calculation, we find the most contribution
to the global $\Lambda$ polarization comes from the $\boldsymbol{\varpi}_{S}$
term in Eq.~(\ref{eq:spin_components}) rather than the $\boldsymbol{\varpi}_{T}$
term. Similar result can also be found in Ref. \cite{Karpenko:2016jyx}.
Therefore the global polarization in Eq.~(\ref{eq:polarization})
can be approximated as
\begin{equation}
P=\frac{1}{N}\sum_{i=1}^{N}\frac{C}{2}\varpi_{zx}(x_{i}),\label{eq:polarization_zx}
\end{equation}
where $\varpi_{zx}(x_{i})$ is the $y$-component of $\boldsymbol{\varpi}_{S}$
in Eq.~(\ref{eq:vorticity_components}) at the space-time point of
the $i$-th $\Lambda$, and the coefficient $C$ encapsulates the
contribution from the ratio $E_{p}/m$ in Eq.~(\ref{eq:spin_components})
and the Lorentz boost correction from $\mathbf{S}$ to $\mathbf{S}^{*}$
in Eq.~(\ref{eq:spin_rest_frame}). In the non-relativistic limit,
$\Lambda$'s energy-momentum $(E_{p},\mathbf{p})$ tends to $(m,0)$
which leads to $C=1$, so one can treat the coefficient $C$ as a
relativistic correction. By comparing the global polarization calculated
from Eqs.~(\ref{eq:spin_components}-\ref{eq:polarization}) with
the one from Eq.~(\ref{eq:polarization_zx}), we find $C$ is around
1 which is not sensitive to the collisional energy. Then the energy
behavior of the polarization is approximately proportional to the
rest part of Eq.~(\ref{eq:polarization_zx}), which we can rewrite
in an integration form
\begin{equation}
P\propto\int d^{4}x\,f_{\Lambda}(x)\varpi_{zx}(x),\label{eq:polarization_int}
\end{equation}
where we have omitted the coefficient $C/2$, and $f_{\Lambda}(x)$
is the space-time distribution of $\Lambda$ at hadronization. One
can see clearly from Eq.~(\ref{eq:polarization_int}) that the global
polarization is jointly determined by the space-time distribution
of $\Lambda$ and the thermal vorticity field $\varpi_{zx}$.

\begin{figure}
\begin{centering}
\includegraphics[width=0.8\columnwidth]{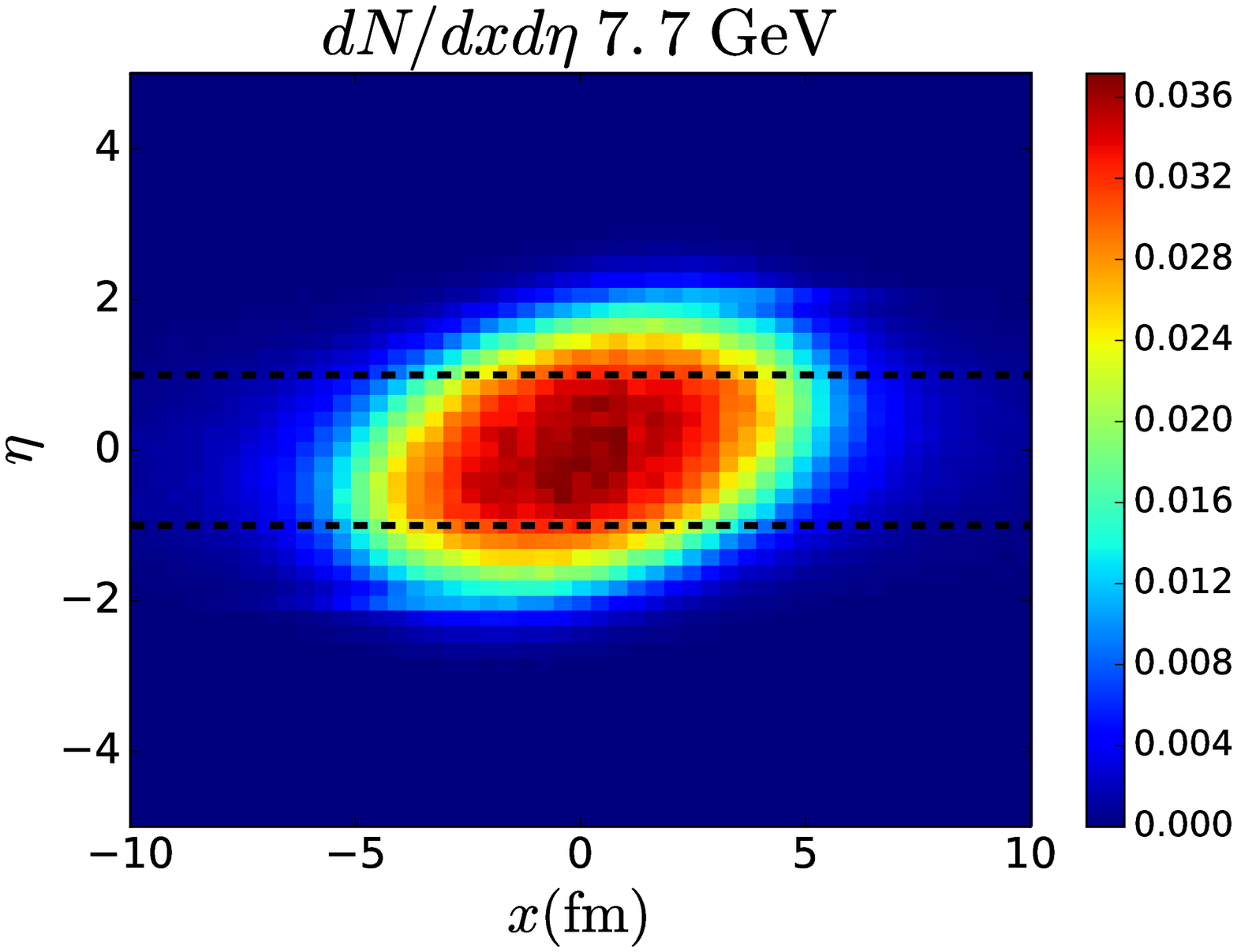}
\par\end{centering}
\begin{centering}
\includegraphics[width=0.8\columnwidth]{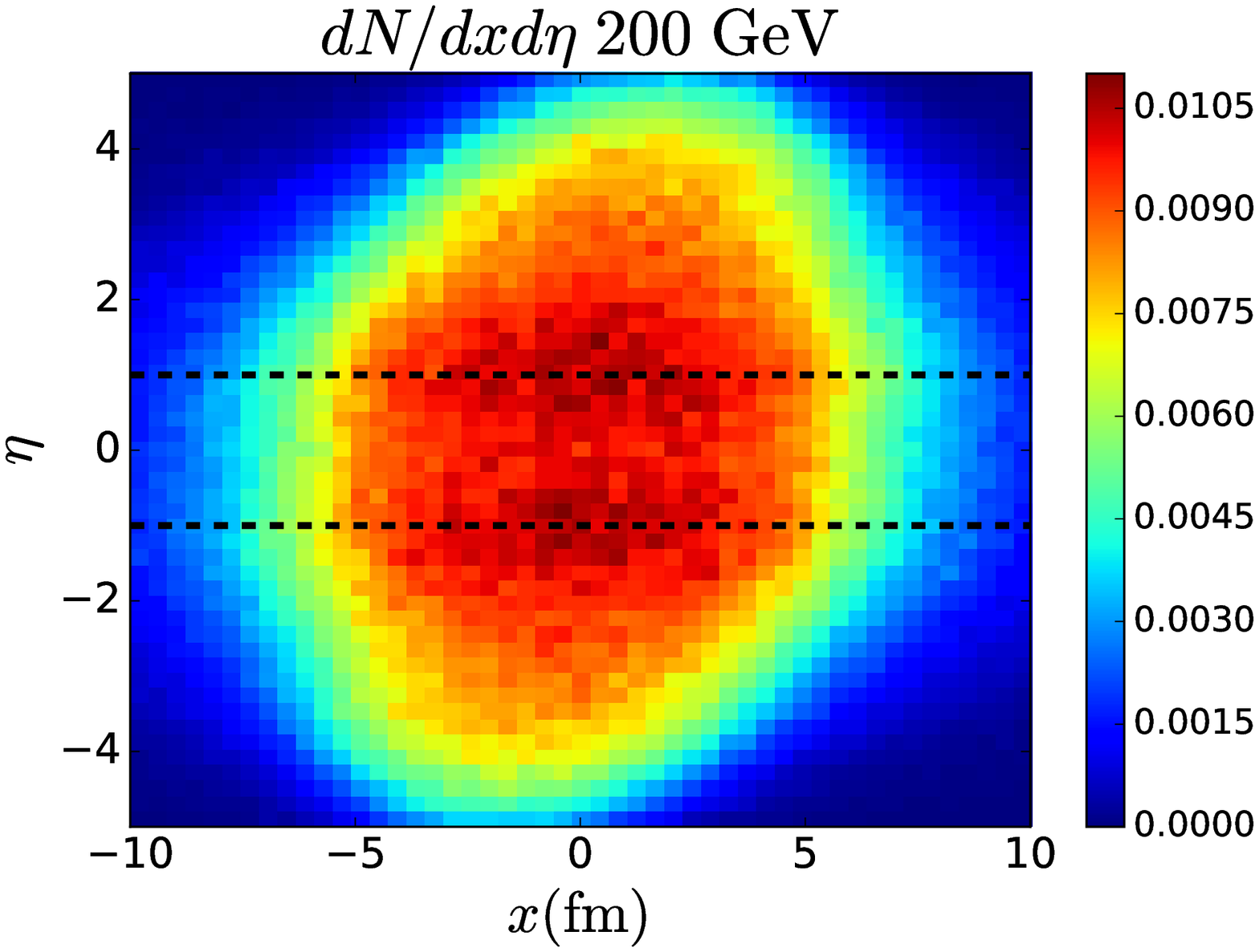}
\par\end{centering}
\caption{\label{fig:lambda-distribution} The distribution of the $\Lambda$'s
production position integrated over $t$ and $y$ as a function of
$x$ and $\eta$ at $\sqrt{s_{NN}}=7.7$ GeV (upper panel) and 200
GeV (lower panel). The midrapidity region $|\eta|<1$ is between the
black dashed lines.}
\end{figure}

\begin{figure}
\begin{centering}
\includegraphics[width=0.8\columnwidth]{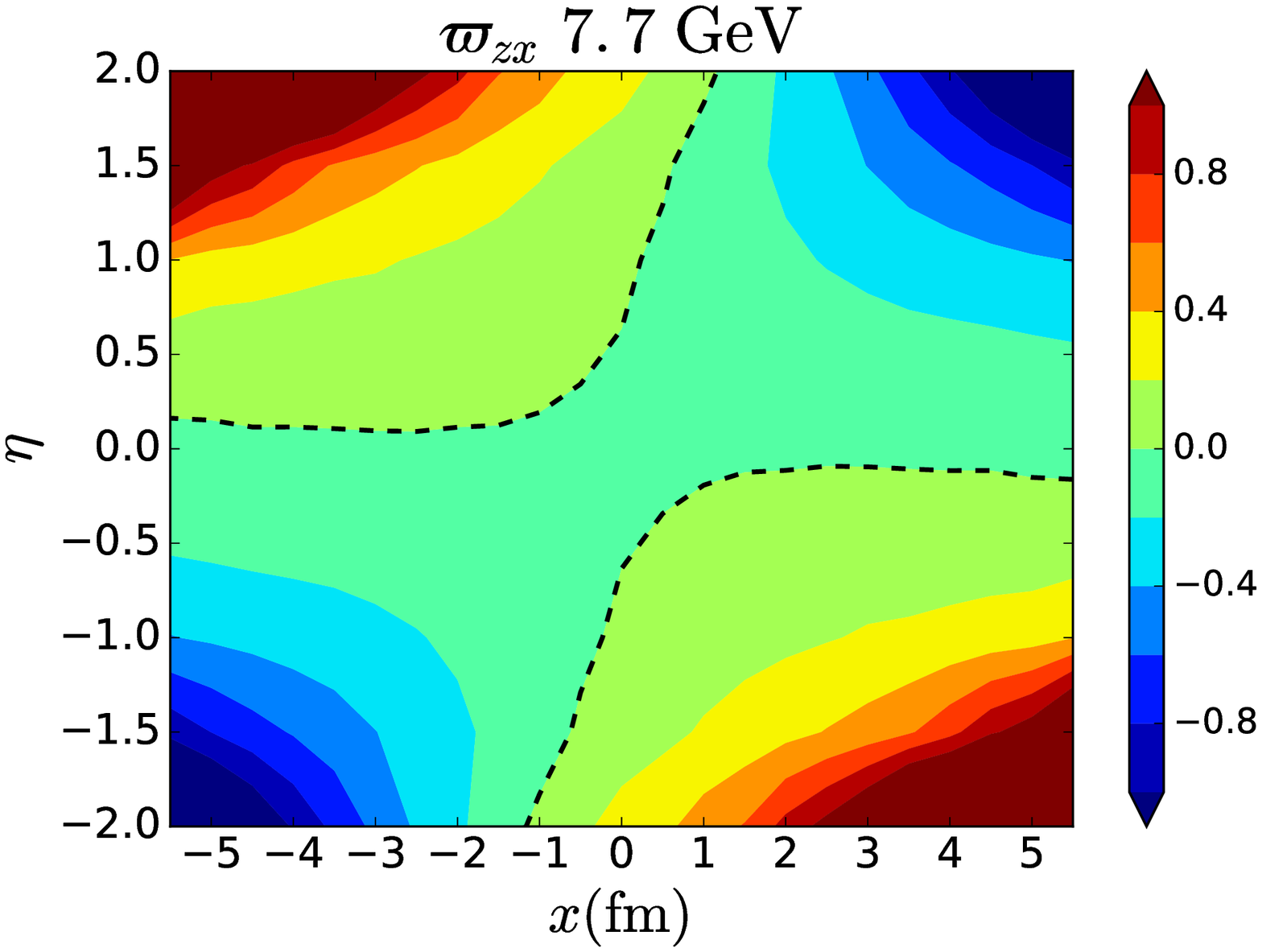}
\par\end{centering}
\begin{centering}
\includegraphics[width=0.8\columnwidth]{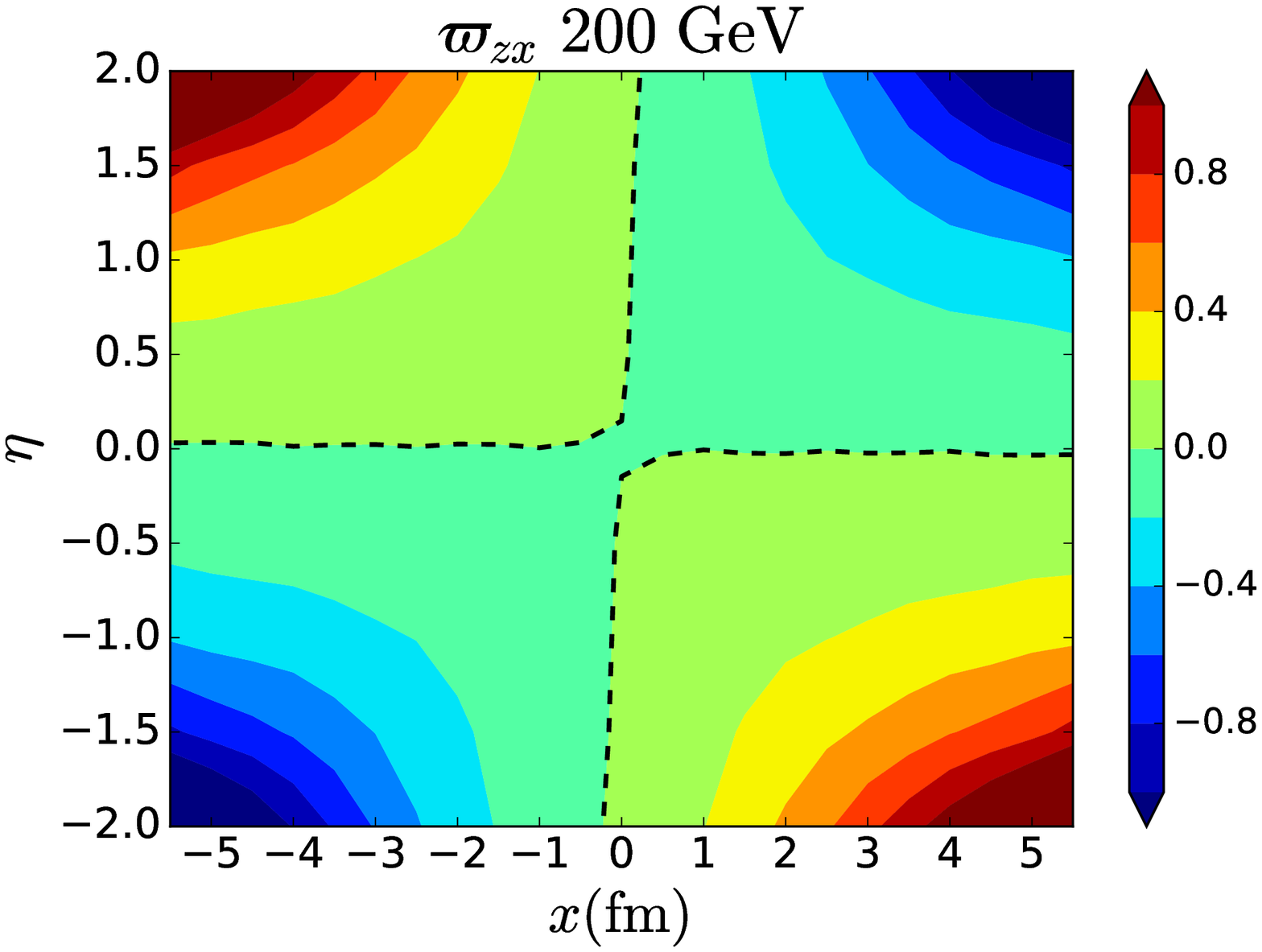}
\par\end{centering}
\caption{\label{fig:omega} The thermal vorticity $\varpi_{zx}$ on the reaction
plane ($y=0$) at $t=5\ \text{fm}/c$ at $\sqrt{s_{NN}}=7.7$ GeV
(upper panel) and 200 GeV (lower panel). The black dashed lines
represent the contour where $\varpi_{zx}=0$. }
\end{figure}

In the following, we investigate the energy dependence of $f_{\Lambda}$
and $\varpi_{zx}$ and study how they combine to determine the energy
behavior of the polarization. We show $f_{\Lambda}$ and $\varpi_{zx}$
in Fig.~\ref{fig:lambda-distribution} and \ref{fig:omega} separately
for $\sqrt{s_{NN}}=7.7$ GeV and 200 GeV. The results at other BES
energies between these two energies can be regarded as some kind of
interpolation between them. We also select $b=7$ fm for illustration.

Figure \ref{fig:lambda-distribution} shows the distribution of the
$\Lambda$'s production position integrated over $t$ and $y$, so
it is a function of $x$ and the space-time rapidity $\eta$. We see
that $f_{\Lambda}$ has a sidewards tilt, namely more $\Lambda$ are
produced in the upper-right and lower-left region due to an asymmetric
matter density distribution in off-central collisions. In the midrapidity
region $|\eta|<1$ (between the black dashed lines in Fig.~\ref{fig:lambda-distribution})
that we are interested in, $f_{\Lambda}$ still shows a tilt at 7.7
GeV, but it is almost symmetric in both $x$ and $\eta$ at 200 GeV.
The latter is the result of a broader rapidity range over which the
fireball extends at higher energies so that the tilt can only be observed
at large rapidity.

Figure \ref{fig:omega} shows the spatial distribution of $\varpi_{zx}$
on the reaction plane ($y=0$), in which we take $t=5\ \text{fm}/c$
as an example for illustration. Here $\varpi_{zx}$ is not weighted
by the particle number or energy density. We can see the thermal vorticity
field $\varpi_{zx}$ shows a quadrupole structure, i.e., it has opposite
signs on different sides of each axis. Similar patterns are also found
in other models or simulations \cite{Becattini:2015ska,Teryaev:2015gxa,Jiang:2016woz,Ivanov:2017dff}.
At 200 GeV the thermal vorticity field $\varpi_{zx}$ is nearly a
perfectly odd function of both $x$ and $\eta$. This structure can
be understood by the radial flow of the system, in which the transverse
velocity $v_{x}$ is an odd function of $x$ but an even function
of $\eta$ \cite{Jiang:2016woz}. At 7.7 GeV the thermal vorticity
field $\varpi_{zx}$ is not an odd function as evidenced by the fact
that $\varpi_{zx}$ is non-vanishing in the central region $x\simeq0$
and $\eta\simeq0$. We also see in Fig.~\ref{fig:omega} that $\varpi_{zx}$
has the same magnitude at 7.7 and 200 GeV. For the time evolution
of $\varpi_{zx}$, we have checked the magnitude of $\varpi_{zx}$
decays with time and the decay rate is not sensitive to the collisional
energy. Therefore $\varpi_{zx}$ at different energies are always
at the same magnitude during the time evolution. We also checked that
the pattern of $\varpi_{zx}$ in spatial distribution does not change
with time at each collisional energy.

Given the space-time distribution of $f_{\Lambda}$ and $\varpi_{zx}$
at 7.7 and 200 GeV, we now study how they combine to give the global
polarization. We first look at the effect of their spatial distribution.
We know that $\varpi_{zx}$ is negative in the upper-right and lower-left
region and leads to a $\Lambda$ polarization along the $-y$ direction,
and $\varpi_{zx}$ is positive in the upper-left and lower-right region
and gives a $\Lambda$ polarization along the $+y$ direction. When
taking the average, these opposite polarizations cancel each other.
Therefore the global polarization depends on how many $\Lambda$ hyperons
are produced in the positive and negative-vorticity region.

At 200 GeV, as shown in Figs.~\ref{fig:lambda-distribution} and
\ref{fig:omega}, $\varpi_{zx}$ ($f_{\Lambda}$) is nearly a perfectly
odd (even) function in both $x$ and $\eta$ in the midrapidity region.
There is almost an equal number of $\Lambda$ hyperons produced in
the positive and the negative-vorticity region. Therefore the global
$\Lambda$ polarization is almost vanishing at 200 GeV. At 7.7 GeV,
there are more $\Lambda$ hyperons produced in the negative-vorticity
region because: (1) $\varpi_{zx}$ is negative in the central region
($x\simeq0$ and $\eta\simeq0$); (2) the tilt shape of $f_{\Lambda}$
leads more $\Lambda$s produced in the upper-right and lower-left
region than the upper-left and lower-right region. As the result,
the global $\Lambda$ polarization at 7.7 GeV is significantly non-zero.

\begin{figure}
\begin{centering}
\includegraphics[width=0.5\textwidth]{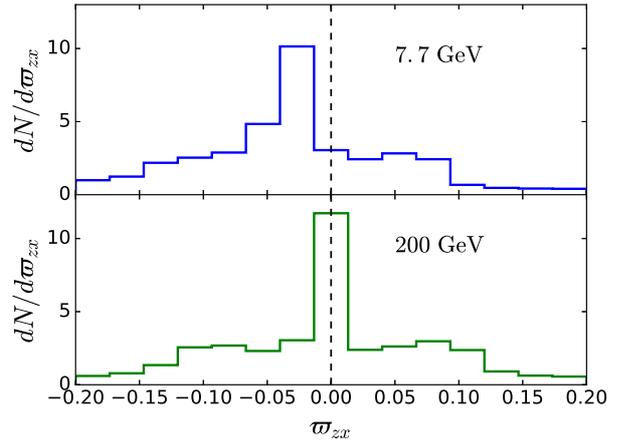}
\par\end{centering}
\caption{\label{fig:lambda distribution-1}The distribution of the number of
$\Lambda$ hyperons $dN/d\varpi_{zx}$}
\end{figure}

The above argument is supported by Fig.~\ref{fig:lambda distribution-1}
where we discretize the values of $\varpi_{zx}$ into several bins
and count the number of $\Lambda$ hyperons produced in the region
with the specific $\varpi_{zx}$ value. The figure clearly shows more
$\Lambda$ hyperons are produced in the negative-vorticity region
at 7.7 GeV, while an almost equal number of $\Lambda$ hyperons is
produced in both the positive and the negative-vorticity region at
200 GeV.

Beside the spatial distribution, the global polarization is also related
to when $\Lambda$ hyperons are produced. Due to the lower temperature
of the fireball, the mean $\Lambda$ production time at 7.7 GeV is
earlier than that at 200 GeV. As $\varpi_{zx}$ decays with time,
when $\Lambda$ hyperons are produced at 200 GeV the magnitude of
$\varpi_{zx}$ is smaller than that at 7.7 GeV. This effect also contributes
to the energy behavior of the global polarization.

Both angular momentum and global polarization are related to the vorticity.
The angular momentum is an integral effect of vorticity weighted by
the moment of inertia over the volume of fireball,
\begin{equation}
\mathbf{J}=\int d^{3}x\,I(x)\boldsymbol{\omega}(x),
\end{equation}
where $I(x)$ is the moment of inertia density of fireball and $\boldsymbol{\omega}=\nabla\times\mathbf{v}/2$
is the non-relativistic vorticity. The exact form of $I(x)$ in fireball
is not clear. A well motivated assumption is $I(x)$ being proportional
to the particle number or energy density, see the discussions in \cite{Jiang:2016woz}.
The total amount of the moment of inertia increases with the collisional
energy, and so does the total angular momentum of fireball. Such behavior
is opposite to the global $\Lambda$ polarization. However, in the
midrapidity region of fireball, the angular momentum should decrease
as the collisional energy increases, because $I(x)$ is nearly symmetric
in the positive and the negative-vorticity region at high energy,
just like $f_{\Lambda}(x)$. In this way, the energy dependence of
$\Lambda$ polarization can be understood by the smaller angular momentum
deposited at midrapidity for higher collisional energies. We also
note that what happens in the midrapidity region at high energy is
quite similar to the situation in central collisions ($b=0$), in
which $f_{\Lambda}(x)$ and $I(x)$ ($\varpi_{zx}$ and $\omega_{y}$)
are exactly even (odd) functions of $x$ and $\eta$, therefore even
though non-zero local vorticity is generated, the total angular momentum
and global $\Lambda$ polarization are vanishing after taking the
integral (or average).

\section{Summary}

\label{sec:summary}In this paper we calculated the global $\Lambda$
polarization in Au+Au collisions at BES energies $\sqrt{s_{NN}}=7.7-200$
GeV with the AMPT model. With the feed-down $\Lambda$ correction
from resonance decays, the magnitude of the global polarization increases
from about 0.2\% to 2.1\% as the collisional energy decreases from
200 to 7.7 GeV which agrees with experimental measurements at STAR
\cite{STAR:2017ckg} within the error-bars.

To explain this energy behavior, we extracted the dominant contribution
to the global polarization as Eq.~(\ref{eq:polarization_int}). The
global polarization is jointly determined by the space-time distribution
of $\Lambda$ and the thermal vorticity field. The larger global polarization
at lower collisional energies is due to (1) more $\Lambda$s are produced
in the negative-vorticity region at lower energies due to larger sidewards
tilt and slower expansion, and (2) earlier $\Lambda$s production
at lower energies which means the magnitude of vorticity does not
decay too much.

\vphantom{}
\begin{acknowledgments}
The authors thank Yin Jiang, Jinfeng Liao, Zi-Wei Lin, Michael A.
Lisa, Dirk H. Rischke, Zebo Tang, and Zhangbu Xu for helpful discussions.
We also thank the anonymous referee for useful comments. HL, QW and
XLX are supported in part by the Major State Basic Research Development
Program (973 Program) in China under Grant No. 2015CB856902 and 2014CB845402
and by the National Natural Science Foundation of China (NSFC) under
Grant No. 11535012. LGP acknowledges funding through the Helmholtz
Young Investigator Group VH-NG-822 from the Helmholtz Association
and the GSI Helmholtzzentrum für Schwerionenforschung (GSI).
\end{acknowledgments}

\bibliographystyle{apsrev4-1}

\end{document}